\documentclass[a4paper,11pt]{article}
\usepackage{pos}
\usepackage{enumitem}
\usepackage{makecell}
\usepackage{multirow}
\usepackage{multicol}

\renewcommand{\logo}{\relax}

\title{Improving Air Shower Simulations by Tuning Pythia\,8/Angantyr with Accelerator Data}

\author*[a]{Michael Windau}
\author[b]{Chloé Gaudu}
\author[b]{Karl-Heinz Kampert}
\author[a]{Kevin Kröninger}

\affiliation[a]{Technische Universität Dortmund, Otto-Hahn-Straße 4a, 44227 Dortmund, Germany}
\affiliation[b]{Bergische Universität Wuppertal, Gaußstraße 20, 42119 Wuppertal, Germany}

\emailAdd{michael.windau@tu-dortmund.de}
\emailAdd{gaudu@uni-wuppertal.de}
\emailAdd{kampert@uni-wuppertal.de}
\emailAdd{kevin.kroeninger@tu-dortmund.de}

\abstract{We present a combined analysis of the Pythia\,8 event generator using accelerator data and evaluate its impact on air shower observables.\\
Reliable simulations with event generators are essential for particle physics analyses, achievable  through advanced tuning to experimental data. Pythia\,8 has emerged as a promising high-energy interaction model for cosmic ray air shower simulations, offering well-documented parameter settings and a user-friendly interface to enable automatic tuning efforts. Using data from collider and fixed-target experiments, we first derive tunes for each domain separately, before tuning both domains simultaneously. To achieve this, we define a core set of observables and quantify their dependence on selected parameters.
The tuning efforts are based on gradient descent and Bayesian methods, the latter providing a full uncertainty propagation of the parameters to the observables.

Results for the impact of a combined analysis for the Pythia\,8/Angantyr event generator on air shower observables, such as particle densities at ground level and energy deposit profiles, are presented.}

\FullConference{39th International Cosmic Ray Conference (ICRC2025)\\
 15–24 July 2025\\
Geneva, Switzerland\\}

\begin{document}
\maketitle

\section*{Introduction}

Monte Carlo event generators are important tools for describing experiments in which particles collide at relativistic energies. They are an integral part of the full simulation chain, which includes modeling the source of the colliding particles, their interactions, the propagation and decay of the final-state particles, and the detector response. While there are event generators designed for specific processes, there are also general-purpose generators that aim at describing a broader range of physical processes and applications. Their applications include collider and fixed-target experiments, as well as air shower observations. Examples of general-purpose event generators include Pythia, Herwig, and SHERPA. They are based on models that describe different aspects of particle interactions, such as hadronization, parton showering, and fragmentation. These models typically have free parameters that must be adjusted — or \emph{tuned} — to experimental data. This tuning is often based on data sets similar to the intended application, e.g., event generators are often tuned using data from other colliders, such as the Large Electron-Positron Collider (LEP), the Large Hadron Collider (LHC), and Tevatron.

In the long run, our goal is to improve the description of air shower data by developing a coherent model that incorporates data from collider and fixed-target experiments. Because of its well-documented parameter descriptions and user-friendly interface, we use the Pythia\,8 event generator~\cite{Bierlich:2022pfr} as a start to describe parton-level scattering in all applications. 
Previous efforts to tune Pythia\,8 include the Monash tune~\cite{Skands:2014pea}, which improved the modeling of underlying event and multiparton interactions over a broad range of collider energies. A dedicated Forward Physics tune~\cite{Fieg:2023kld} has enhanced the description of particle production in the very forward region at the LHC. Building on these developments, ongoing work aims to tune Pythia\,8 within CORSIKA\,8 for air shower physics, focusing on improving muon production by incorporating fixed-target data to address the so-called \emph{Muon Puzzle} observed in cosmic-ray experiments~\cite{Gaudu:2024mkp}.
For this study, a subset of the free parameters in Pythia 8 using selected data sets from
LEP and LHC experiments, as well as data from fixed-target experiments. 
Rather than pursuing a global tune that includes all available datasets, our goal is to investigate the interplay between different datasets and test the compatibility of tuned parameter values for datasets from various sources. Finally, we will examine the effect on air shower simulations performed using CORSIKA\,8, a modern, modular framework for simulating extensive air showers \cite{Engel:2018akg, Alameddine:2024cyd, Riehn:ICRC2025}.
It is written in C++ which offers improved flexibility, maintainability, and extensibility for incorporating new physics models and components. Recent efforts have focused integrating the hadronic event generator Pythia\,8/Angantyr~\cite{Bierlich:2022pfr, Bierlich:2018xfw, Bierlich:2021poz} into CORSIKA\,8. This allows for detailed studies of high-energy interactions and facilitates model tuning against both collider and air shower data~\cite{Gaudu:2024rsq, Gaudu:2025K1, Gaudu:2025qhv}.

In Section~\ref{sec:data}, we provide a brief summary of the software/tools as well as the chosen parameters and datasets, and the observables used within the scope of this study. The Bayesian tuning setup is detailed in Section~\ref{sec:tuning}. We also illustrate how the observables are parameterized with respect to the model parameters. The tuning results are described and discussed in detail in Section~\ref{sec:results}. Air showers using the tuned results are simulated and discussed in Section~\ref{sec:air_shower_results}. We conclude the study with a summary and outlook in Section~\ref{sec:conclusions}.

\section{Parameter and data set selection}
\label{sec:data}


Although several tunes are available for Pythia that adjust a large set of parameters using data from the LEP, LHC, SPS and the Tevatron (e.g.\ the Monash tune~\cite{Skands:2014pea}), we restrict our initial study to three parameters that are expected to influence the description of collider and fixed-target data:
\begin{itemize}[noitemsep, partopsep=0pt, parsep=0pt]
    \item \texttt{MultipartonInteractions:pT0Ref} is a reference transverse momentum ($p_T$) essential for the regularization of QCD cross sections at small $p_T$. It is expected to impact hadronic collision processes.
    \item \texttt{StringZ:aLund} is the $a$ parameter in the Lund fragmentation model~\cite{Andersson:1983ia}. It is expected to impact all processes involving hadrons in the final state.
    \item \texttt{StringPT:sigma} is the spread of the transverse momentum of the produced hadrons. It is expected to impact all processes involving hadrons in the final state.
\end{itemize}

We selected a set of observables affected by these parameters and then considered the corresponding datasets from various experiments. These include charged particle multiplicities (L3 and ATLAS experiments), distributions of the log of the scaled momentum, $x_p=2|\vec{p}|/E_\mathrm{cm}$ (L3 and ALEPH experiments), and transverse momentum distributions of charged particles (ATLAS and NA22 experiments). The data are accessed via an interface to RIVET~\cite{Bierlich:2019rhm}. RIVET was developed to support the development, validation, and tuning of event generators. It uses C++ plugins to encapsulate analysis logic in a reproducible, modular format. The output is stored in YODA~\cite{Buckley:2023xqh}, which is a human-readable format that can be populated with data from sources such as HEPData~\cite{Maguire:2017ypu}. A RIVET plugin is used to determine how to transform raw event output. It does this by selecting particles, applying filters, and binning observables in order to match the structure of experimental measurements. \texttt{Pythia8Rivet} provides a direct event stream from Pythia\,8 to RIVET, whereas other generators generally rely on full HepMC~\cite{Dobbs:2001ck} files as an intermediary. Table~\ref{tab:rivet_plugin} summarizes the datasets used and categorizes them as "LEP", "LHC" or "fixed-target" dataset collections.

\begin{table}[h]
    \centering
    \begin{tabular}{|c|c|c|c|c|} \cline{2-5}
        \multicolumn{1}{c|}{} & RIVET plugin & Data & Observable & Ref \\ \hline
        \multirow{3}{*}{\raisebox{\totalheight}{LEP}} & L3\_2004\_I652683 &  \makecell{d59-x01-y01 \\ d65-x01-y01} & \makecell{$P(N_\mathrm{ch})$ \\ d$\sigma$/d$\zeta_p$} & \cite{L3:2004cdh} \\ \cline{2-5}
         & ALEPH\_1996\_I428072 & d17-x01-y01 & d$\sigma$/d$\zeta_p$ & \cite{ALEPH:1996oqp} \\ \hline
        \multirow{2}{*}{\raisebox{\totalheight}{LHC}} & ATLAS\_2010\_I882098 & \makecell{d10-x01-y01 \\ d17-x01-y01} & \makecell{d$\sigma$/d$\eta$d$p_\perp$ \\ d$\sigma$/d$N_\mathrm{ch}$} & \cite{ATLAS:2010jvh} \\ \hline
         fixed-target & EHS\_1988\_I265504 & d06-x01-y01 & d$\sigma$/d$p_T^2$ & \cite{EHSNA22:1988fqa} \\ \hline
    \end{tabular}
    \caption{RIVET plugins used in the combined analysis.}
    \label{tab:rivet_plugin}
\end{table}
\vspace{-0.5cm}
\section{Tuning procedure}
\label{sec:tuning} 


In this study, we employ a tuning procedure based on the methodology developed in~\cite{LaCagnina:2023yvi}. It is based on Bayesian inference and uses Markov chain Monte Carlo methods to calculate the posterior probability densities of the parameters under study. The implementation relies on the \texttt{BAT.jl} ~\cite{Schulz:2020ebm} and \texttt{EFTFitter.j} ~\cite{Castro:2016jjv} JULIA packages. The tuning procedure involves constructing a parametric model $\vec{f}$ representing the response of the event generator and inferring on the free parameters of the model, $\vec{\lambda}$, using data $\vec{D}$. 
The components of $\vec{f}$ and $\vec{D}$ represent predictions and corresponding measurements for individual observables, such as the predicted bin content of a distribution. The vectors are sized according to the number of predictions and measurements, $n$. The data are assumed to have Gaussian uncertainties and potential correlations, which are summarized in a covariance matrix $M$. The resulting likelihood takes the following form
\begin{equation}
\mathrm{ln} L(\vec{D}|\vec{\lambda}) = -\frac{1}{2}[\vec{D}-\vec{f}(\vec{\lambda})]^T M^{-1} [\vec{D}-\vec{f}(\vec{\lambda})] \, .
\end{equation}
We choose uniform distributions for the priors of all parameters. The resulting posterior probability distributions are sampled using the Metropolis Hastings algorithm with ten parallel chains and $10^5$ sampling steps per chain. We report the global mode as the resulting parameter set from the tuning procedure. We also indicate the smallest 68\% credible intervals from the marginalized distributions for each parameter to illustrate the range of uncertainty.

In addition to providing a set of tuned parameters, this method provides the full posterior distribution $p(\lambda|\vec{D})$, which allows for an evaluation of the uncertainty of individual observables resulting from the tuning: by sampling from the posterior distribution, the posterior distribution for each prediction $f_{i}$, $p(f_i|\vec{D})$ is calculated and the smallest 68\% credible interval is reported as the uncertainty of the prediction. 

A focus has been placed on the construction of the parametric model $\vec{f}(\vec{\lambda})$. While the event generator can be run each time the function is evaluated, this approach is not computationally feasible. Instead, we parameterize the response of the event generator as a function of $\vec{\lambda}$. We define a grid in the parameter space and and generate $xyz$ events for each configuration. These events are then used to calculate the predictions $\vec{f}$. The resulting sets of predictions often exhibit complex features that simple polynomial parameterizations cannot capture. Therefore, the functional form, $\vec{f}(\vec{\lambda})$, is calculated using linear interpolation between neighboring grid points. 
Although the parameterization yields exact values at the support points, the linear interpolation has proven stable and accurate enough for the tuning process.


\section{Tune results}
\label{sec:results}

\begin{figure}[h]
  \noindent
  \begin{minipage}{0.5\textwidth}
    \centering
    \includegraphics[width=\linewidth]{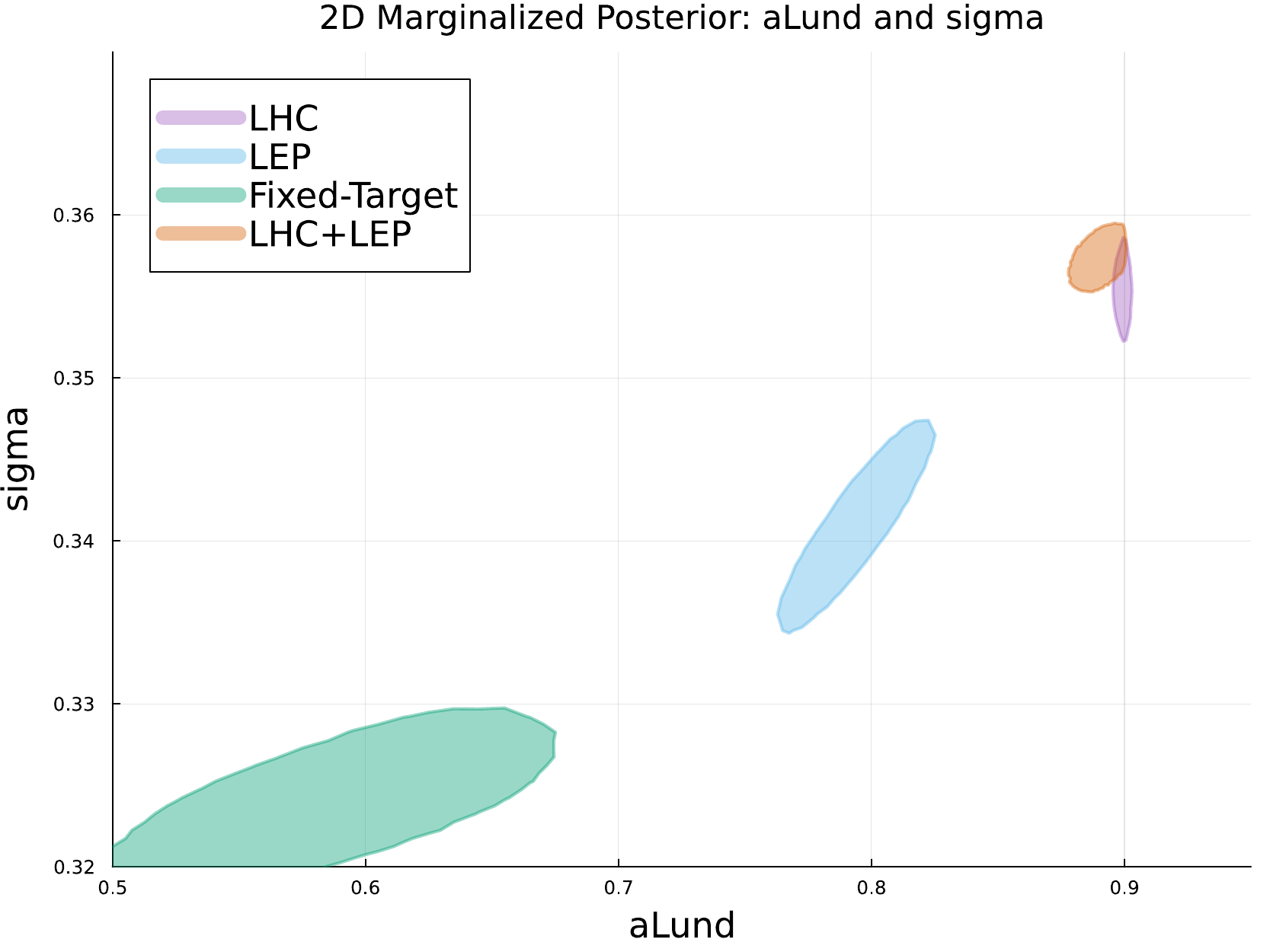}
  \end{minipage}%
  \hfill
  \begin{minipage}{0.5\textwidth}
    \centering
    \includegraphics[width=\linewidth]{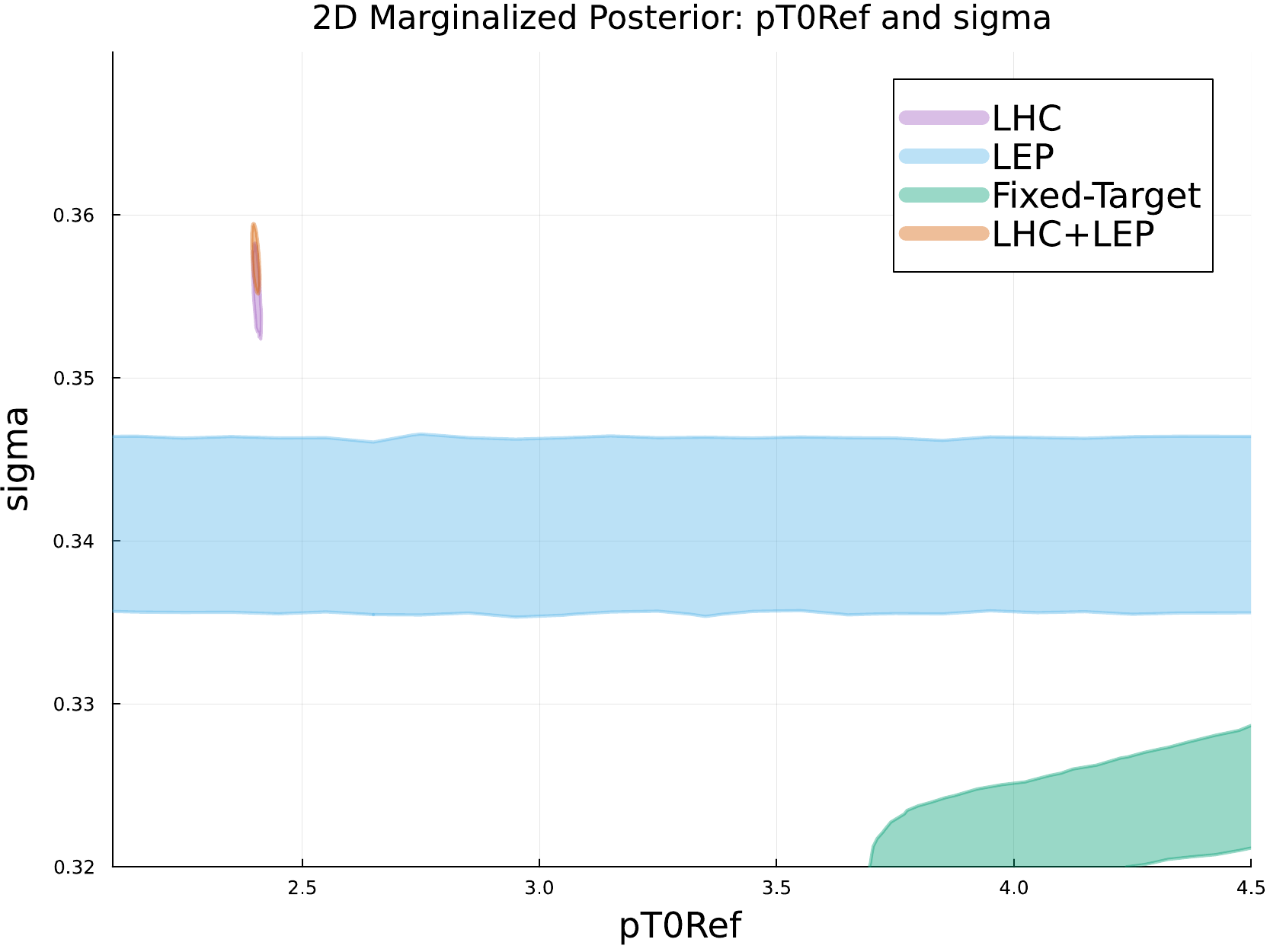}
  \end{minipage}
  \vspace*{-0.2cm}
  \captionof{figure}{%
  Marginalized 2D posteriors of the (\texttt{aLund}, \texttt{sigma}) parameters (left) and the (\texttt{pT0Ref}, \texttt{sigma}) parameters (right) for the different tunes.}
  \label{fig:2DPosteriors}
\end{figure}

We apply the tuning procedure presented in Section~\ref{sec:tuning} to the LHC, LEP and fixed-target data sets separately. The tuned values obtained from the LHC and LEP data sets are reasonably consistent. However, there is a significant discrepancy in the values obtained from the fixed-target data, particularly for the \texttt{pT0Ref} parameter. This discrepancy complicates a combined tune across all selected experiments. Specifically, forcing a single tune would result in a \texttt{pT0Ref} value incompatible with each individual dataset. 
Consequently, we opted for two tunes: one for LEP+LHC observables and one for the fixed-target observable. 
The results of the tuning process are presented in Table~\ref{tab:p8_parameters}. 
Furthermore, Figure~\ref{fig:2DPosteriors} shows marginalized 2D posterior distributions for the LHC, LEP, LHC+LEP, and fixed-target tunes. This confirms our observation that, while the fixed-target posterior is in a similar range in \texttt{sigma}, the offset in \texttt{aLund} to the LHC, and more significantly in \texttt{pT0Ref} compared to the LEP and LHC posteriors, is too large to justify a single combined tune at the cost of reduced agreement with data.

\begin{figure}[h]
  \noindent
  \vspace{-0.5cm}
  \begin{minipage}{0.5\textwidth}
    \centering
    \includegraphics[width=\linewidth]{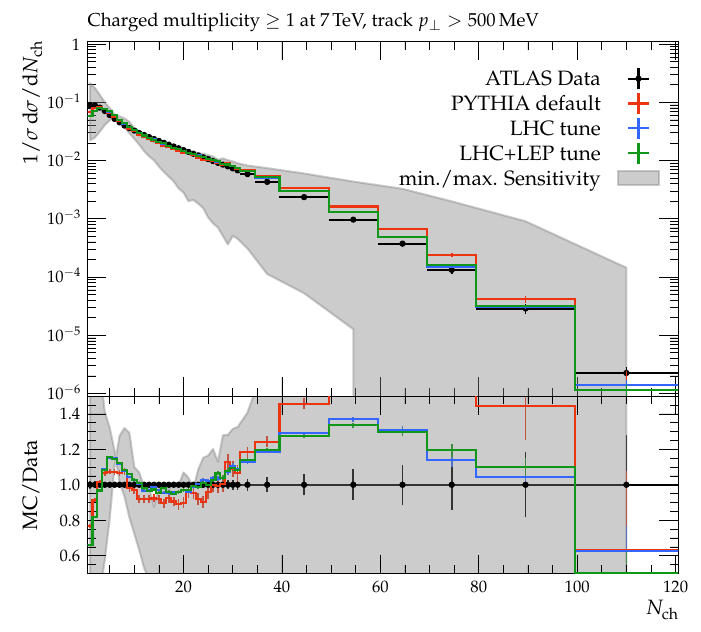}
  \end{minipage}%
  \hfill
  \begin{minipage}{0.5\textwidth}
    \centering
    \includegraphics[width=\linewidth]{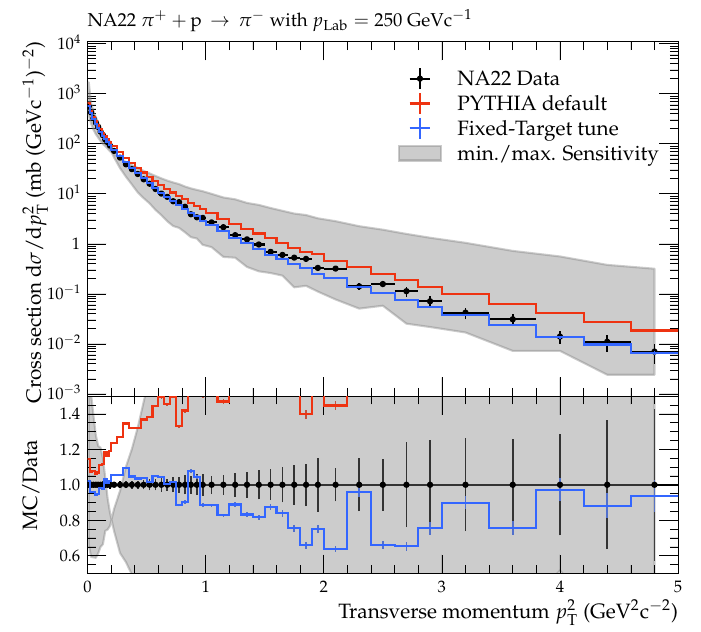}
  \end{minipage}
  \hfill
  \begin{minipage}{0.5\textwidth}
    \centering
    \includegraphics[width=\linewidth]{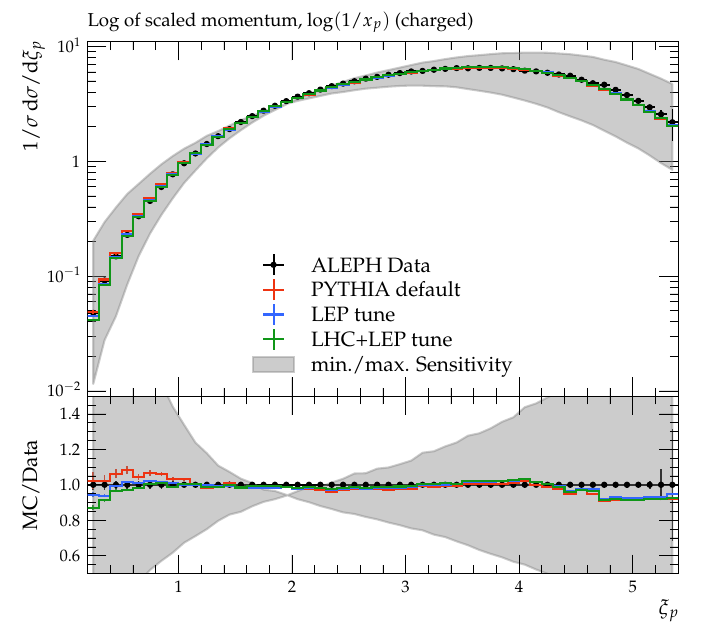}
  \end{minipage}
  \hfill
  \begin{minipage}{0.5\textwidth}
    \centering
    \includegraphics[width=\linewidth]{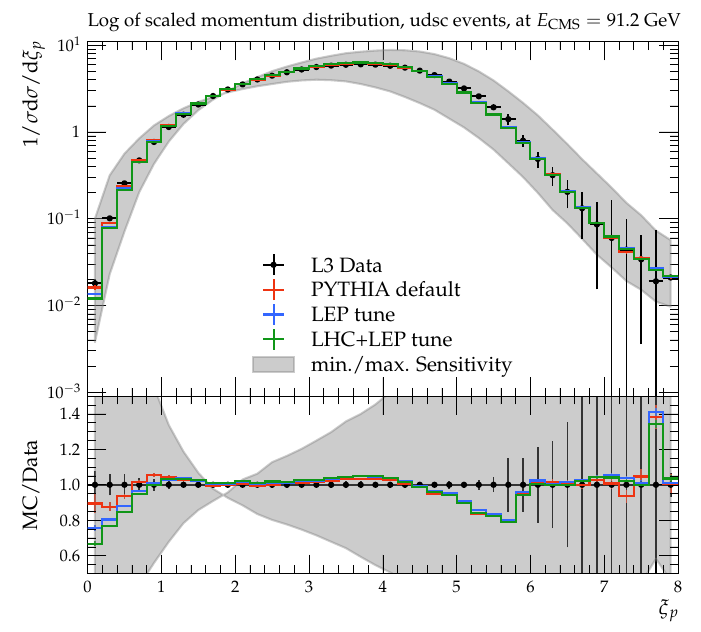}
  \end{minipage}
  \vspace*{-0.2cm}
  \captionof{figure}{%
  Distribution of the charged multiplicity observable from ATLAS ~\cite{ATLAS:2010jvh} (top left), the transverse momentum of charged particles from NA22 ~\cite{EHSNA22:1988fqa} (top right), the log of the scaled momentum from ALEPH ~\cite{ALEPH:1996oqp} (bottom left) and L3 ~\cite{L3:2004cdh} (bottom right) for the data and the tuned and default event generator samples. The gray band indicates the range between the minimum and maximum allowed parameter values.}
  \label{fig:observables}
  \vspace{-0.5cm}
\end{figure}

Figure~\ref{fig:observables} highlights the impact of the different tunes by comparing event generator predictions to data for four of the selected observables. For the NA22 fixed-target observable, the fixed-target tune notably reduces the reduced $\chi^2$-value from 100.7 (default) to 5.7. For the charged multiplicity observable from ATLAS, the LEP+LHC tune also improves the results, reducing the reduced $\chi^2$-value from 230.3 to 99.4. For the log of scaled momentum observable from L3, the tune reduces the reduced $\chi^2$-value from 12.3 to 10.0. However, not all observables improve; the observable from ALEPH deteriorates slightly under the LEP+LHC tune, with the reduced $\chi^2$-value increasing from 4.0 to 5.6.


\begin{table}[h]
    \centering
    \begin{tabular}{|c|c|c|c|} \cline{2-4}
        \multicolumn{1}{c|}{} & \texttt{MultipartonInteractions:pT0Ref} & \texttt{StringZ:aLund}  & \texttt{StringPT:sigma} \\ \hline 
        min & 0.5 & 0 & 0 \\ \hline
        default & $2.28$ & $0.68$  & $0.335$ \\ \hline
        max & 10 & 2 & 1 \\ \hline \hline
        LHC+LEP & $2.404\pm0.004$ & $0.890\pm0.005$ & $0.357\pm0.001$  \\ \hline
        fixed-target & $4.780\pm0.574$\footnotemark  & $0.640\pm0.090$ & $0.326\pm0.004$ \\ \hline 
    \end{tabular}
    \caption{Parameter values of Pythia\,8/Angantyr in its default and modified configurations.} 
    \label{tab:p8_parameters}
\end{table}
\footnotetext{The \texttt{MultipartonInteractions:pT0Ref} value for the fixed-target tune lies at the upper limit of the chosen parameter range.}
\vspace{-0.5cm}
\section{Impact on air shower observables}
\label{sec:air_shower_results}

To investigate the impact of Pythia tunes based on either LHC+LEP or fixed-target measurements, we ran CORSIKA\,8 air shower simulations. In these simulations, we replaced the high-energy hadronic interaction model with our custom-tuned version of Pythia. For this work, we simulated a total of 1000 air showers induced by protons at $10^{17}$\,eV with a zenith angle of $30^\circ$, using Pythia\,8.315/Angantyr with both the default and modified parameter sets displayed in Table~\ref{tab:p8_parameters}. 
The particle tracking energy thresholds were set at 10\,MeV for electrons, positrons, and photons, and 300\,MeV for hadrons and muons. The transition from the FLUKA low-energy model  \cite{Ballarini:2024uxz} to Pythia\,8/Angantyr occurs at 100 GeV.

\begin{figure}[h]
  \noindent
  \vspace{-0.4cm}
  \begin{minipage}{0.49\textwidth}
    \centering
    \hspace*{-0.7cm}
    \includegraphics[width=1.1\linewidth]{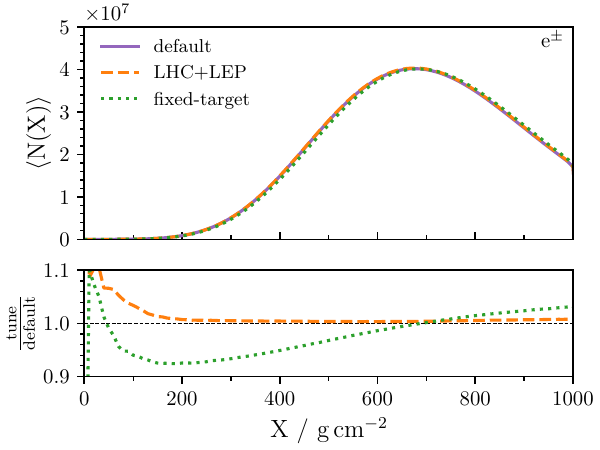}
  \end{minipage}%
  \hfill
  \begin{minipage}{0.49\textwidth}
    \centering
    \hspace*{-0.2cm}
    \includegraphics[width=1.1\linewidth]{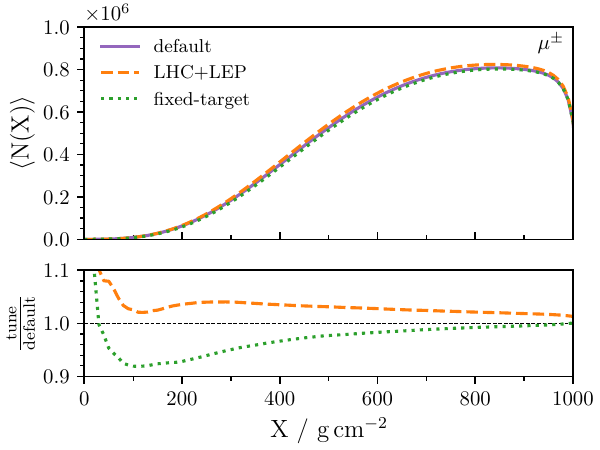}
  \end{minipage}
  \captionof{figure}{%
  Average longitudinal shower profiles of electrons and positrons (left) and muons (right) for proton-induced air showers at $E=10^{17}$\,eV and
  $\theta = 30^\circ$ simulated with CORSIKA\,8+Pythia\,8.315/Angantyr for its default and modified parameters configurations.}
  \label{fig:long_mu}
  \vspace{-0.4cm}
\end{figure}
Figure~\ref{fig:long_mu} shows that the average longitudinal profiles of electrons and positrons are nearly identical for the default and LHC+LEP tunes. In contrast, the fixed-target tune exhibits a significantly reduced number of electrons and positrons up to about $X\sim700$\,g\,cm$^{-2}$, followed by higher number than the default beyond this depth. Regarding the muon component: the LHC+LEP tune produces more muons than the default throughout the shower development. In contrast, the fixed-target tune produces fewer muons overall, yet converges to that of the default near the observation level.

\begin{figure}[h]
  \noindent
  \begin{minipage}{0.49\textwidth}
    \centering
    \includegraphics[width=\linewidth]{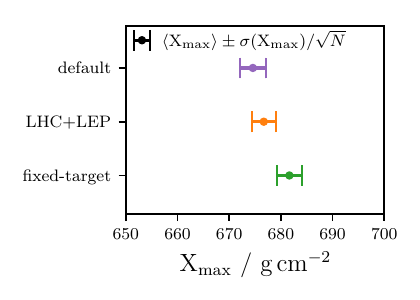}
  \end{minipage}%
  \hfill
  \begin{minipage}{0.49\textwidth}
    \centering
    \includegraphics[width=\linewidth]{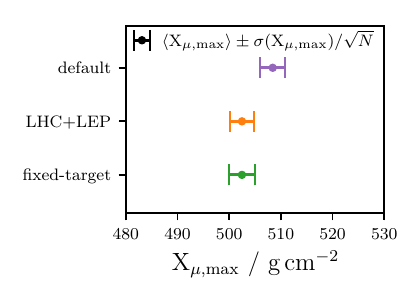}
  \end{minipage}
  \vspace*{-0.6cm}
  \captionof{figure}{%
  Average depth of the shower maximum $X_\mathrm{max}$ (left) and average depth of the muonic shower maximum $X_{\mu,\mathrm{max}}$ (right) for 1000 proton-induced air showers at $E=10^{17}$\,eV and
  $\theta = 30^\circ$ simulated with CORSIKA\,8+Pythia\,8.315/Angantyr for its default and modified parameters configurations.}
  \label{fig:Xmax_Xmumax}
  \vspace{-0.5cm}
\end{figure}

Figure~\ref{fig:Xmax_Xmumax} shows the average electromagnetic and muonic shower maxima, which differ by roughly 2\% between the tunes. For the electromagnetic maximum, the default and LHC+LEP tune largely overlap within uncertainties, indicating only a minor shift, whereas the fixed-target tune lies at a deeper $X_\mathrm{max}$ with partial overlap with LHC+LEP, resulting in, on average, deeper showers. 
In contrast, the muonic maximum decreases from the default to both tunes, reflecting an earlier development of the muon component.

\begin{figure}[]
  \noindent
  \begin{minipage}{0.49\textwidth}
    \centering
    \includegraphics[width=\linewidth]{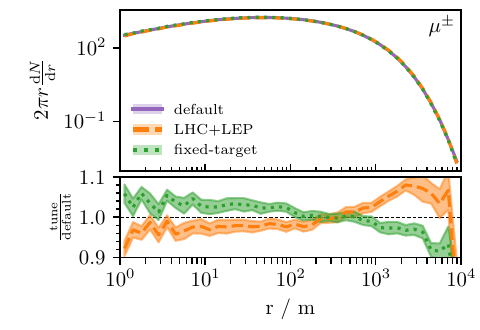}
  \end{minipage}%
  \hfill
  \begin{minipage}{0.49\textwidth}
    \centering
    \includegraphics[width=\linewidth]{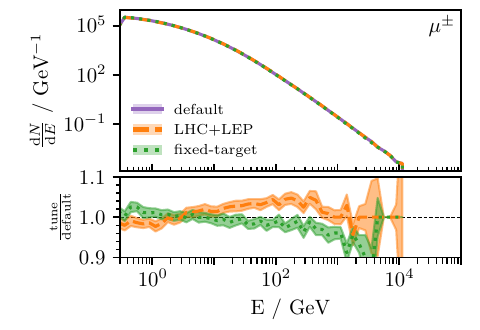}
  \end{minipage}
  \vspace*{-0.4cm}
  \captionof{figure}{%
  Median lateral distributions (left) and median energy spectra (right) of muons at Auger height for proton primaries at $E=10^{17}$\,eV and
  $\theta = 30^\circ$ simulated with CORSIKA\,8+Pythia\,8.315/Angantyr for its default and modified parameters configurations. The shaded bands indicate the interquartile range (25\%–75\%).}
  \label{fig:ldf_espect}
  \vspace{-0.5cm}
\end{figure}
From the lateral distribution of muons shown in Figure~\ref{fig:ldf_espect}, one observes that both the fixed-target and LHC+LEP tunes exhibit deviations at the few-percent level near the shower core. The fixed-target tune leads to an increased number of muons by up to approximately 5\%, whereas the LHC+LEP tune results in a slight deficit compared to the default settings. However, this trend reverses around 300 m from the shower core: the LHC+LEP tune then yields up to 10\% more muons than the default, while the fixed-target tune produces about 10\% fewer muons at the same radial distance. In the muon energy spectrum, we observe deviations of up to 5\% for both tunes relative to the default settings, with a characteristic switching behavior occurring around $\sim$30-40 GeV. Below this threshold, the LHC+LEP tune yields fewer low-energy muons, while the fixed-target tune produces slightly more. Above the threshold, this trend reverses: the LHC+LEP tune results in an excess of high-energy muons, whereas the fixed-target tune shows a deficit. At the highest muon energies, statistical fluctuations dominate, limiting us in drawing firm conclusions.
\section{Conclusions}
\label{sec:conclusions}

We have presented a tune of three parameters of the Pythia\,8 event generator that are relevant to physics processes at LEP, the LHC, and fixed-target experiments. We observe significant differences in the resulting parameters between the collider data and fixed target data, in particular for \texttt{pt0Ref}. As a consequence, we extract two separate sets of tuned parameter values and use these two setting for the simulation of air showers. We find differences of up to 10\% in the average longitudinal shower profiles of electrons and muons and differences of up to 2\% in the average depth of the shower maximum and the muonic shower maximum. Future studies will include further data, observables and parameters, and they will focus on resolving or confirming the resulting discrepancies. 


\setlength{\bibsep}{0pt}
\setlength{\itemsep}{0pt}
\setlength{\parskip}{0pt}
{\footnotesize
\begin{multicols}{2}

\end{multicols}
}

\noindent
\normalsize
\section*{Acknowledgements}
This work was supported by the Deutsche Forschungsgemeinschaft (DFG, German Research Foundation) through the Collaborative Research Center SFB1491 “Cosmic Interacting Matters - From Source to Signal”.  The computations were partially carried out on the PLEIADES cluster at the University of Wuppertal, which was supported by the Deutsche Forschungsgemeinschaft (DFG, grant No. INST 218/78-1 FUGG) and the Bundesministerium für Bildung und Forschung (BMBF).

\end{document}